\documentstyle[prd,
preprint,
aps]{revtex}
\begin{document}
\draft
\preprint{\mbox{KIAS-P98022} }

\title{Comment on the possible electron neutrino excess
       in the Super-Kamiokande atmospheric neutrino experiment}
\author{C. W. Kim }
\address{Department of Physics and Astronomy, 
The Johns Hopkins University,
Baltimore, MD21218, USA}
\address{   Korea Institute for Advanced Study,
                        207-43 Cheongryangri-dong,
                        Dongdaemun-gu,
                        Seoul 130-012, 
                        Korea 
            }
\author{U. W. Lee}             
\address{
         Department of Physics,
         Mokpo National University,
         Muan-gun, Chonnam 534-729,
         Korea
        }
\address{Email address: cwkim@kias.kaist.ac.kr, 
leeuw@chungkye.mokpo.ac.kr}
\maketitle

\begin{abstract}
We investigate the implications of 
a possible excess of the electron neutrino events
in the Super-Kamiokande atmospheric neutrino experiment.
The excess, if real,
leads to constraints on the 
$\nu_{\mu} \leftrightarrow \nu_e$, $\nu_{\mu}\leftrightarrow
\nu_{\tau}$ and 
$\nu_{e} \leftrightarrow \nu_{\tau}$  
oscillation parameters. 
Specifically, the excess implies that $\nu_e - \nu_\mu$ 
and $\nu_\mu - \nu_\tau$ mixings are large whereas 
$\nu_e - \nu_\tau$ mixing is small.
We show that the electron neutrino excess
favors the large mixing angle MSW solution 
of the solar neutrino problem. 
\end{abstract}

\pacs{PACS number(s): 14.60Lm, 14.60.Pq, 25.65}

%
%\narrowtext
%
%\begin{multicols}{2}
%

Recently, 
the Super-Kamiokande Collaboration has
reported the evidence for the neutrino oscillation 
from the atmospheric neutrino experiment\cite{SUPERK}.
It has been remarkably well demonstrated that the data are consistent
with the 
$\nu_\mu \leftrightarrow \nu_\tau$ oscillations
with $\sin^2 2\theta > 0.82$ and
$ 
5 \times 10^{-4} 
<
\Delta m^2
<
6 \times 10^{-3}
{{\mathrm  eV}}^2
$
at 90 \% confidence level.
However, 
the data appear to imply that there exists an excess of 
the number of detected electrons 
compared to the calculated number 
by Monte Carlo method\cite{MCATM1,MCATM2,MCATM3}
for the up-going low energy neutrinos.
The Fig.1 in Ref.\cite{SUPERK}
is the plot of the up-down asymmetry 
for the e-like and $\mu$-like events.
One can easily see the electron event excess
compared to the Monte Carlo calculation
for the low energy electrons.
The Fig.3 in Ref.\cite{SUPERK}
is the plot of the zenith angle distribution
of the e-like and $\mu$-like events for the sub-GeV and multi-GeV
data sets.
For the e-like events 
there are several data points which do not agree
with 
the Monte Carlo calculation.
In particular, a data point for the e-like events for
$ p < 0.4 $ GeV with 
$-1.0 \leq \cos \theta \leq -0.6$ 
is more than several $\sigma$'s above the expected number.
The Fig.4 in Ref.\cite{SUPERK}
provides another hint of the electron excess. 
Even though the data are well fitted with the dashed line
({\em i. e.} consistent with $\nu_\mu \leftrightarrow \nu_\tau$ 
and no $\nu_e$ oscillations),
the largest $L/E$ data of the e-like events
has the largest deviation from the dashed line.
Therefore, the three figures in Ref.\cite{SUPERK} 
consistently suggest the existence of the electron neutrino excess 
for the up-going low energy neutrinos.
Although there exist large uncertainties in the
number of detected neutrinos
calculated by Monte Carlo method\cite{MCATM1,MCATM2,MCATM3}, 
it is worthwhile to investigate  possible implications of this
electron neutrino excess.
From the study of the excess of electrons
in the framework of three-generation 
neutrino oscillations,
we have obtained the constraints(or information) 
on the neutrino oscillation parameters.

For the sake of simplicity, 
we assume the CP conservation 
in the lepton sector
and the relation
\begin{equation}
P(\nu_\alpha\to\nu_\beta)
=
P(\nu_\beta\to\nu_\alpha)
=
P_{\alpha  \beta}
 \ ,
\end{equation}
where $\alpha , \  \beta$ represent $e$ , $\mu$ and $\tau$.
Furthermore, 
we use the notation $N_\alpha^{MC}$ 
for the expected  number of $\nu_\alpha$ induced events
calculated by the Monte Carlo method
for the Super-Kamiokande experiment 
and
the $N_\alpha^{DATA}$ 
for the observed number of the $\nu_\alpha$ induced events
in the Super-Kamiokande experiment as published in Ref.[1], 
respectively. 

With these conventions,
the numbers of the detected e-like and $\mu$-like events  
are given by 
\begin{eqnarray}
N_e^{DATA} 
%&=& 
%       N_e^{MC} \left[ 1 - P_{e \mu} - P_{e \tau} \right]
%       + N_\mu^{MC} P_{e \mu } \nonumber \\
&=& N_e^{MC} 
          +     \left[ N_\mu^{MC} - N_e^{MC} \right] P_{e \mu}
          -  N_e^{MC} P_{e  \tau}
\ ,
\label{NNUE}
\\
N_\mu^{DATA} 
%&=& 
%       N_\mu^{MC} \left[ 1 - P_{\mu e} - P_{\mu  \tau} \right]
%       + N_e^{MC} P_{\mu  e} \nonumber  \\
&=& N_\mu^{MC} 
  -     \left[ N_\mu^{MC} - N_e^{MC} \right] P_{e \mu} 
  -  N_\mu^{MC} P_{\mu \tau}
\label{NNMU}
\ .
\end{eqnarray}
 $P_{\alpha \beta}$ is the 
$\nu_\alpha \leftrightarrow \nu_\beta$ 
transition probability and 
$P_{\alpha\beta} \propto \sin^2 2 \theta_{\alpha\beta}$
where $\theta_{\alpha\beta}$ represents an ``effective''
$\nu_\alpha  - \nu_\beta$ mixing angle. (We caution the reader that
this angle is not one of the CKM mixing angles.)
As pointed out already, 
the Super-Kamiokande data show the existence of 
an excess of the up-going low energy electrons.
Therefore, from  Eq.(\ref{NNUE}) 
we have the following inequality
\begin{equation}
\left[ N_\mu^{MC} - N_e^{MC} \right] \ P_{e \mu}
-  N_e^{MC} 
\ P_{e \tau} \
> 0    . 
\label{PEMUPETAU}
\end{equation}
If the 
excess of the e-like events is real,
we obtain a very strong constraint 
on the neutrino oscillation probabilities
$P_{e \mu}$ and $P_{e \tau}$ 
in the parameter regions 
where the electron neutrino excess has been detected.
Equation (\ref{PEMUPETAU}) can be rewritten as
\begin{equation}
P_{e\mu} 
>
\frac{1}
     {  N_\mu^{MC} / N_e^{MC}   - 1 }
\ P_{e \tau} ~ .
\label{PEMUI}
\end{equation}
This inequality is a function of 
$  N_\mu^{MC} / N_e^{MC} $.
Although 
$N_\mu^{MC}$ and $N_e^{MC}$
themselves have large uncertainties,
their ratio is known to the accuracy
of about  5 \%
even among several different flux estimates.
(The uncertainty in each calculation is much smaller 
than 5 \%.)
Therefore, 
Eq.(\ref{PEMUI}) is insensitive 
to theoretical errors in  each neutrino flux.
Let us examine the sub-GeV atmospheric neutrino data 
used in the recent super-Kamiokande atmospheric neutrino
experiment. From Fig.1 of Ref.\cite{SUPERK} we obtain
the expected numbers of the electrons and muons
as
%the constraint on $P_{e \mu}$ and $P_{e \tau}$ as
\begin{equation}
\frac{N_\mu^{MC}}{N_e^{MC}} 
= 
1.224 \pm 0.007 ,
\label{EQU1224}
\end{equation}
for the events in the bin with $ -1< \cos \theta <-0.6 $ and $p < 0.4 $ GeV. Therefore, the following analysis is based on these events, i.e. All the limits on and values of the transition probabilities are valid for those nutrinos although the results on the mixing angles 
are of course valid in general.
We emphasize here that in obtaining Eq.(\ref{EQU1224}),
$N_{\mu}^{MC}$ and $N_e^{MC}$ must be shifted simultaneously
up or down within the error bars
rather than taking, {\em e.g.}, 
an upper value of $N_{\mu}^{MC}$ 
and a lower value of $N_e^{MC}$
in order to obtain an upper bound on the ratio
$N_{\mu}^{MC} / N_e^{MC}$ and so on.
(Also, it is to be noted that the number in 
Eq.(\ref{EQU1224}) is based on the calculation in Ref.\cite{MCATM1}
and can be slightly different for different calculations
within errors of about 5 \%.)
Hence, with Eq.(\ref{PEMUI}), we find
\begin{equation}
P_{e \mu} 
>
4.35 P_{e \tau}
\label{CGSUB}
\end{equation}
for the up-going low energy neutrinos.
This equation suggests that since $P_{e\mu} \leq 1$
\begin{equation}
P_{e\tau} < 0.23
\end{equation}
in the parameter regions where the electron neutrino excess was 
detected in the Super-Kamiokande experiment.

%\mbox{}

Now, let us calculate some bounds on $P_{e\mu}$
for the up-going low energy neutrinos.
Since $P_{e\tau} \geq 0$, 
we have, from Eq.(\ref{NNUE}),
the relation 
\begin{equation}
P_{e\mu }
\geq
\frac{N_e^{DATA}/N_e^{MC} - 1}
       {  N_\mu^{MC} / N_e^{MC}  - 1}
\ .
\label{PEMUIEQ}
\end{equation}
As mentioned already, although the ratio 
$(N_\mu^{MC} / N_e^{MC})$
is known to a fairly good accuracy,
the ratio
$(N_e^{DATA} / N_e^{MC})$
has the uncertainties inherent in $N_e^{MC}$.
If we simply use the ratio of the two central values of
$N_e^{DATA}$ and $N_e^{MC}$, 
which is $1.45$,
we obtain from Eq.(\ref{PEMUIEQ}),
$P_{e\mu} > 2$, 
which is unphysical.
Even after taking all the uncertainties as quoted 
in Fig.(3) of Ref.\cite{SUPERK},
we still obtain $P_{e\mu} > 1$,
suggesting either too high $N_e^{DATA}$ or
too low $N_e^{MC}$(or both).
This problem persists even if a somewhat higher value of
$N_e^{MC}$ as given in Ref.\cite{MCATM2}
is used.
Therefore, as was done in Ref.\cite{SUPERK} to find the best fit,
we raise simultaneously
$N_e^{MC}$ and $N_\mu^{MC}$ by $15.8 \% $.
Then taking all the uncertainties into account, we find
\begin{equation}
P_{e\mu} \geq 0.51
\ .
\label{PEMU051}
\end{equation}
Defining ``effective'' mixing angle $\theta_{e\mu}$ and ``effective''
$\Delta m_{e\mu}^2$ 
in the form of two generation neutrinos as
\begin{equation}
P_{e\mu} =
\sin^2 2 \theta_{e\mu} \sin^2 
\left(
1.27 \frac{\Delta m_{e\mu}^2}{{\mathrm eV}^2} 
     \frac{L / {\mathrm km}}{E / {\mathrm GeV}} 
\right)
\ ,
\end{equation}
we obtain from Eq.(\ref{PEMU051})
\begin{equation}
\sin^2 2 \theta_{e\mu} \geq 0.51
\ .
\end{equation}
One can also consider an extreme rescaling by 20 \%
as was done in Ref.\cite{FOGLI}.
In this case we find
\begin{equation}
\sin^2 2 \theta_{e\mu} \geq 0.33 .
\end{equation}
Thus, our result that
$\sin^2 2 \theta_{e\mu}$ is ``significantly''
larger than $10^{-3} \sim 10^{-2}$ 
(which is the value of the small mixing angle MSW solution)
is rather robust.
We, therefore, conclude that
the large mixing angle MSW solution of the solar neutrino 
problem is favored over the small mixing angle solution.

It is interesting to note 
that the electron excess is prominent only for the 
up-going low energy $\nu_e$-induced events.
This means that these neutrinos with
$p \simeq E < 0.4 \ {\mathrm GeV}$ 
have the longest distance to travel through the earth.
They correspond to a data point at slightly above
$(L/{\mathrm km})/(E/{\mathrm GeV}) = 10^4$ 
in Fig.4 in Ref.\cite{SUPERK},
which is indeed elevated.
This implies that for other electron neutrinos,
the values of
$\Delta m_{e\mu}^2 L/E$ are expected to be less than
$\pi / 2$
(or 
$\sin^2 
\left(
1.27 \frac{\Delta m_{e\mu}^2}{{\mathrm eV}^2}
                   \frac{L / {\mathrm km}}
                       {E / {\mathrm GeV}} 
\right) 
<< 
1$)
so that $\nu_e \leftrightarrow \nu_\mu$ 
oscillations would not take place.
For this situation to be realized, 
$1.27 \Delta m_{e\mu}^2 L/E$
must become roughly
$\pi /2$ only when $L \simeq 2 R_E$ and 
$E < 0.4 \ {\mathrm GeV}$
(we take $0.3 \ {\mathrm GeV}$ 
for events with with $ E < 0.4 \ {\mathrm GeV}$).
We then find
\begin{equation}
\Delta m_{e\mu}^2 \simeq 10^{-5} \ {\mathrm eV}^2
\ ,
\end{equation}
which is indeed consistent with the value of
$\Delta m_{e\mu}^2$ obtained as 
the MSW solution of the solar neutrino problem\cite{SOL,66},
further rendering a support to our conclusion.
Furthermore, 
the oscillation parameters are consistent with
the constraint obtained from the
CHOOZ experiment \cite{CHOOZ}
because the ``effective'' $\Delta m_{e\mu}^2$ 
is well below the forbidden region of  the CHOOZ experiment.

Next we consider $P_{\mu\tau}$.  
From unitarity and Eq.(\ref{PEMU051}), 
we find 
\begin{equation}
P_{\mu\tau} \leq 0.49
\ . 
\label{PMUT049}
\end{equation}
This does not necessarily mean that
$\sin^2 2 \theta_{\mu\tau} \leq 0.49$,
which disagrees with
$\sin^2 2 \theta_{\mu\tau} \simeq 1$
implied by the two generation fit in Ref.\cite{SUPERK}.
If we adapt 
$\Delta m_{\mu\tau}^2 \simeq 10^{-3} {\mathrm eV}^2$,
which is the result given in
Ref.\cite{SUPERK},
$\Delta m_{\mu\tau}^2 L/E$ 
becomes much larger than the oscillation length
for $L \simeq 2 R_E$ and 
$E < 0.4 \ {\mathrm GeV}$.
Thus we expect the oscillation to be in the rapid oscillation stage
with the average
$
\left< 
\sin^2 \left(
1.27 \frac{\Delta m_{\mu\tau}^2}{{\mathrm eV}^2} 
     \frac{L / {\mathrm km}}{E / {\mathrm GeV}} 
        \right) 
\right>
= \frac{1}{2}
$,
so that Eq.(\ref{PMUT049}) gives
\begin{equation}
\sin^2 2 \theta_{\mu\tau} \leq 0.98
\ ,
\end{equation}
which is not inconsistent with the result,
$\sin^2 2 \theta_{\mu\tau} \simeq 1$, in Ref.\cite{SUPERK}

%%%%%
In conclusion,
we need the combined analysis of 
$\nu_e \leftrightarrow \nu_\mu$,
$\nu_e \leftrightarrow \nu_\tau$ and 
$\nu_\mu \leftrightarrow \nu_\tau$ oscillations
to explain the atmospheric neutrino data\cite{THREENU}.
The small $\Delta m_{e\mu}^2 $ of order 
$10^{-5 } {\mathrm eV}^2$ with a large 
$\nu_e - \nu_\mu$ mixing angle
provides an explanation for the excess of 
the up-going low energy e-like events.
Therefore,
{\em the large mixing angle MSW solution of the solar neutrino problem
is favored over the small mixing angle solution 
(which is statistically favored)
if the excess of the electron neutrino
in the Super-Kamiokande atmospheric neutrino experiment
turns out to be real}. 
It remains to be seen with great interest
whether the electron excess persists or not 
in the future experiment.
A large $\nu_\mu - \nu_\tau$ mixing angle with 
$\Delta m_{\mu\tau} \simeq 10^{-3} {\mathrm eV}^2$
is also necessary to explain the depletion of the $\mu$-like
events.
The nonzero value of $P_{e\tau}$ 
gives the depletion of the e-like events.
Since we do not see any depletion of the e-like events, 
we can conclude that 
$P_{e\tau} \simeq  0$,
consistent with Eq.(8).
Our findings are in agreement with 
 the conclusion  favoring the large mixing angle MSW
solution of the solar problem in Ref.\cite{GEORGI} and the last two
papers in Ref[9].
A detail numerical analysis with proper statistical considerations
of our study will be presented elsewhere.

%\mbox{}

\section*{Acknowledgments}

U. W. Lee gratefully thanks KIAS for the hospitality during his visiting
and this work is supported in part by
Basic Science Research Institute of Mokpo National University.
The authors would like thank Carlo Giunti for very helpful comments
and suggestions.


\references
\bibitem{SUPERK} Super-Kamiokande Collaboration, Y. Fukuda, 
                        {\em et al.},
                        Phys. Rev. Lett. 81, 1562(1998).
\bibitem{MCATM1} M. Honda, {\em et. al.}, Phys. Lett. B248, 193(1990);
                                 M. Honda, {\em et. al.}, 
								 Phys. Rev. D52, 4985(1995).
\bibitem{MCATM2} G. Barr, T. K. Gaisser and Todor Stanev,
                        Phys. Rev. D39, 3532(1989); 
                        V. Agrawal, T. K. Gaisser, Paolo Lipari 
						and Todor Stanev,
                        {\em ibid.}, D53, 1314(1996)
\bibitem{MCATM3} T. K. Gaisser, M. Honda, K. Kasahara, H. Lee, 
                        S. Midorikawa, 
                        V.Naumov, and Todor Stanev,
                        Phys. Rev. D54, 5578(1996).     
\bibitem{SOL} Naoya Hata and Paul Langcker,
                        Phys. Rev. D56, 6107 (1997);
\bibitem{66}  J. N. Bachall, P. I. Krastev and A. Yu. Smirnov,
                        hep-ph/9807216, to be published.
\bibitem{FOGLI} G. L. Fogli, E. Lisi, A. Marrone, and G. Scioscia,
			hep-ph/9808205,
		E. Kh. Akhmedov, A. Dighe,   P.  Lipari, A. Yu. Smirnov,
		hep-ph/9808270.
\bibitem{CHOOZ} CHOOZ Collaboration, M. Apollonio {\em et al.},
                        Phys. Lett. B420, 397 (1998).
\bibitem{THREENU} V. Barger, S. Pakvasa, T. J. Weiler, and K. Whisnant,
				hep-ph/9806387;
		Osamu Yasuda, hep-ph/9809205;
		R. N. Mohapatra and S. Nussinov, hep-ph/9809415,
		H. Fritzsch and Z. Z. Xing, Phys. Lett. B372 (1996) 265; 
			{\em ibid} hep-ph/9808272
\bibitem{GEORGI} 	Howard Georgi and S. L. Glashow,
			hep-ph/9808293.
%\end{multicols}
\end{document}